\documentstyle[aps]{revtex}
\begin{document}
%
\title{Beyond the Hubbard-I Solution with a One-Pole Self-Energy at 
Half-Filling within the Moment Approach: Non-Linear Effects.}
\author{J J  Rodr\'{\i}guez - N\'u\~nez}
\address{Universidade Federal de Santa Maria, 
Departamento de F\'{\i}sica - CCNE, 97105-900 
Santa Maria/RS, Brazil \\e-m: jjrn@mfis.ccne.ufsm.br} 
\author{M. Argollo de Menezes}
\address{Instituto de F\'{\i}sica, 
Universidade Federal Fluminense,  
Av.\ Litor\^anea S/N, 
Boa Viagem, 24210-340 Niter\'oi RJ, 
Brazil.} 
\date{\today}
\maketitle

%
%
\begin{abstract}
We have postulated a single pole for the self-energy, 
$\Sigma(\vec{k},\omega)$, looking for the consequences on   
the one-particle Green function, $G(\vec{k},\omega)$ in the 
Hubbard model.  We find that $G(\vec{k},\omega)$ satisfies 
the first two sum rules or moments of Nolting 
(Z. Physik {\bf 255}, 25 (1972)) for any values of the two unknown 
$\vec{k}$ parameters of $\Sigma(\vec{k},\omega)$. In order 
to find these two parameters we have used the third 
and four sum rules of 
Nolting. $G(\vec{k},\omega)$ turns out to be identical to the 
one of Nolting (Z. Physik {\bf 225}, 25 (1972)), which is 
beyond a Hubbard-I solution since satisfies four sum rules. 
With our proposal we have 
been able to obtain an expansion in powers of $U$ for 
the self-energy (here to second order in $U$). We  present numerical 
results at half-filling for 1- the 
static spin susceptibility, $\chi(T)$ vs $T/t$ and 2- the band narrowing parameter, $B(T)$ vs $T/t$. The two-pole 
Ansatz of Nolting for the one-\-par\-ti\-cle Green function is equivalent 
to a single pole Ansatz for the self-e\-ner\-gy which remains the fundamental quantity for more elaborated calculations when, for 
example, lifetime effects are included.
\\
Pacs numbers: 74.20.-Fg, 74.10.-z, 
74.60.-w, 74.72.-h
\end{abstract}

%
%
%
%
	After the discovery of 
the high-$T_c$ materials\cite{Bednorz-Muller}, the study 
of correlations has gained interested due to the fact that there is 
the belief\cite{Anderson} that the normal properties 
of these materials could be explained in the framework of the Hubbard 
model\cite{HubbardI}, since electron correlations are strong, i.e., the on-site 
electron-electron repulsions $U$ are much larger than the energies 
associated with the hybridization of atomic orbitals 
belonging to different atoms\cite{Fulde}.  

	In this paper, we will adopt a given pole structure in the self-energy 
and see its effect on the one-particle Green function. This is done 
with the idea of fleshing out the basic elements implicit in 
the moment approach of Nolting for $G(\vec{k},\omega)$. At the 
same time, we monitor the self-energy which is 
the important quantity where lifetime effects can be incorporated rather easily. We find that by using this approach (one pole in 
$\Sigma(\vec{k},\omega)$ and the sum rules for the spectral functions), the  one-particle Green function goes beyond the Hubbard-I solution, i.e.,  
identically satisfies the first two sum rules and its parameters have 
to be calculated self-consistently. 

    	The model we study is the Hubbard model\cite{HubbardI}
\begin{eqnarray}\label{Ham}
H = t_{\vec{i},\vec{j}}c_{\vec{i}\sigma}^{\dagger}c_{\vec{j}\sigma}
   + \frac{U}{2} n_{\vec{i}\sigma}n_{\vec{i}\bar{\sigma}}   
   - \mu c^{\dagger}_{\vec{i}\sigma}c_{\vec{i}\sigma}~~,
\end{eqnarray}
where $c_{\vec{i}\sigma}^{\dagger}$($c_{\vec{i}\sigma}$) are creation
(annihilation) electron operators with spin $\sigma$. $n_{\vec{i}
\sigma} \equiv c_{\vec{i}\sigma}^{\dagger}c_{\vec{i}\sigma}$. 
$U$ is the local interaction, $\mu$ the chemical potential and we 
work in the grand canonical ensemble. We have adopted Einstein 
convention for repeated indices, i.e., for the $N_s$ sites $\vec{i}$, the 
$z$ nearest-neighbor sites and for spin up and down ($\sigma = \pm 1$). 
$t_{\vec{i},\vec{j}} = -t$, for n.n. and zero otherwise.

	The one-particle 
Green function, $G(\vec{k},\omega)$, is expressed in terms of 
the self-energy, $\Sigma(\vec{k},\omega)$, as
\begin{equation}\label{1PGF}
G(\vec{k};\omega) \equiv  
\frac{1}{\omega - \varepsilon_{\vec{k}} - \Sigma(\vec{k},\omega)} ~~~,
\end{equation}
\noindent where 
$\epsilon(\vec{k}) =-2 t (cos(k_xa) + cos(k_ya))$, 
and $\varepsilon_{\vec{k}} = \epsilon(\vec{k}) - \mu$. 

	We adopt the following Ansatz for $\Sigma(\vec{k},\omega)$:
\begin{equation}\label{ansatz}
\Sigma(\vec{k},\omega) \equiv \rho U +  
\frac{\alpha(\vec{k})}{\omega - \Omega_{\vec{k}}} ~~~.
\end{equation} 

	As $\Sigma(\vec{k},\omega)$ has dimensions of energy, 
the still unknown parameter $\alpha({\vec{k}})$ has dimensions of 
(energy)$^2$. $\alpha(\vec{k})$ is kind of a spectral weight and 
$\Omega_{\vec{k}}$ is the energy spectrum of the self-energy. We will 
calculate $\alpha(\vec{k})$ and $\Omega_{\vec{k}}$.  We have 
included the Hartree shift directly 
in the self-energy (the first term of Eq.(\ref{ansatz})).  
In Eq. (\ref{ansatz}), $\rho$ is 
the carrier concentration per spin orientation, i.e., $\rho = n/2$. We are 
assuming that we are in the paramagnetic phase\cite{magnetic}. The 
physical reason of having chosen a single pole in the self-\-e\-ner\-gy 
is due to the fact that, for intermediate to strong coupling, the 
self-\-e\-ner\-gy gives rise to two energy branches in the 
one-\-par\-ti\-cle and ultimately to the two Hubbard bands. 

	By using Eq. (\ref{ansatz}) into Eq. (\ref{1PGF}), we get 
that the one-particle Green function has two poles. It can be written as
\begin{equation}\label{twopoles}
G(\vec{k},\omega) = \frac{\alpha_1(\vec{k})}{\omega - 
\omega_1(\vec{k})} + \frac{\alpha_2(\vec{k})}{\omega - \omega_2(\vec{k})} ~~~,
\end{equation}
\noindent where 
\begin{eqnarray}\label{solvin}
\omega_1(\vec{k}) &=& \frac{1}{2} \left[ \Omega_{\vec{k}} + \xi_{\vec{k}} + 
\left[\left(\Omega_{\vec{k}} - \xi_{\vec{k}} \right)^2 + 
4\alpha(\vec{k}) \right] ^{1/2} \right] ~~~, ~~~ 
\omega_2(\vec{k}) = \frac{1}{2} \left[ \Omega_{\vec{k}} + \xi_{\vec{k}} - 
\left[\left(\Omega_{\vec{k}} - \xi_{\vec{k}} \right)^2 + 
4\alpha(\vec{k}) \right]^{1/2} \right] ~~~, \nonumber \\ 
\alpha_1(\vec{k}) &=& \frac{\omega_1(\vec{k}) - \Omega_{\vec{k}}}
{\omega_1(\vec{k}) - \omega_2(\vec{k})} ~~~,  ~~~
\alpha_2(\vec{k}) = \frac{\omega_2(\vec{k}) - 
\Omega_{\vec{k}}}{\omega_2(\vec{k}) - \omega_1(\vec{k})} ~~~ .
\end{eqnarray}

	From Eqs. (\ref{solvin}) we inmediately see that
the following sum rules or moments\cite{Nolting} are identically satisfied:
\begin{eqnarray}\label{firstwo}
\alpha_1(\vec{k}) + \alpha_2(\vec{k}) = 1 ~~~ ,  ~~~
\alpha_1(\vec{k}) \omega_1(\vec{k}) + 
\alpha_2(\vec{k}) \omega_2(\vec{k}) = \xi_{\vec{k}} ~~~ ,  ~~~
\xi_{\vec{k}} = \epsilon(\vec{k}) + \rho U ~~~ .
\end{eqnarray}

	Eqs. (\ref{firstwo}) are the 
first two sum rules for the spectral functions of Nolting\cite{Nolting}. 
In order to evaluate $\alpha(\vec{k})$ and $\Omega_{\vec{k}}$, 
we use the next two sum rules of Nolting. This gives:
\begin{eqnarray}\label{nextwo}
\omega_1^2(\vec{k}) \alpha_1(\vec{k}) + \omega_2^2(\vec{k}) \alpha_2(\vec{k}) = 
\xi^2_{\vec{k}} + \alpha(\vec{k}) &=& a_2(\vec{k}) ~~~ , \nonumber \\
\omega_1^3(\vec{k}) 
\alpha_1(\vec{k}) + \omega_2^3(\vec{k}) \alpha_2(\vec{k}) = \xi^3_{\vec{k}} + 
(\Omega_{\vec{k}} + 2\xi_{\vec{k}})\alpha(\vec{k}) &=& a_3(\vec{k}) ~~~ ,
\end{eqnarray}
where $a_2(\vec{k})$,  $a_3(\vec{k})$ are given in 
Ref.\cite{Nolting}-\cite{Micnas-et-al} as
\begin{eqnarray}\label{a2-a3}
a_2(\vec{k}) = 
\epsilon^2(\vec{k}) + 2\rho U \epsilon(\vec{k}) + \rho U^2 ~~~ &,& \nonumber \\
a_3(\vec{k}) = \epsilon^3(\vec{k}) + 3 U \epsilon^2(\vec{k}) (2+\rho) \rho U^2 
\epsilon(\vec{k}) + \rho (1-\rho) U B(\vec{k}) + \rho U^3 ~~~ &,& 
\end{eqnarray}
\noindent 
Temperature enters due to the presence 
of the $B$-\-term. Solving Eqs. (\ref{nextwo}) we find:
\begin{eqnarray}\label{solutions}
\alpha(\vec{k}) = \rho (1 - \rho ) U^2 ~~~ , ~~~ 
\Omega_{\vec{k}} = (1 - \rho) U + B(\vec{k}) ~~~ .
\end{eqnarray}
\indent The narrowing band parameter has to be calculated 
self-consistently and is $\vec{k}$-independent in the spherical 
approximation (See Ref.\cite{Nolting}). Combining Eqs.  (\ref{solvin},\ref{solutions}) we find
\begin{equation}\label{cloNol}
\omega_1(\vec{k}) = \frac{1}{2}\left[H(\vec{k}) + 
[K(\vec{k})]^{1/2}\right] ~~~ ; ~~~ \omega_2(\vec{k}) = 
\frac{1}{2}\left[H(\vec{k}) - [K(\vec{k})]^{1/2}\right] ~~~ , 
\end{equation}
\noindent where
\begin{eqnarray}\label{Noltingregained}
H(\vec{k}) \equiv \epsilon(\vec{k}) U + B(\vec{k}) ~~~ , ~~~ 
K(\vec{k}) \equiv \left(\epsilon(\vec{k}) - U - 
B(\vec{k}) \right)^2 + 4 \rho U \left(\epsilon(\vec{k})  - B(\vec{k}) \right) ~~~ .
\end{eqnarray}
	Eqs. (\ref{Noltingregained}) 
are {\it nothing else} that the solutions given by 
Nolting\cite{Nolting} by solving the four unknown $\alpha_i(\vec{k})$ 
and $\omega_i(\vec{k})$, with $i = 1,2$. We have regained Nolting's solutions 
in a much easier way, starting from the self-energy, while 
Nolting does it from the Green function itself. The self-energy 
has the meaning of being an 
expansion in powers of $U$. In our case, due to the choice of 
a single pole for the self-energy, $\alpha(\vec{k})$ is of second 
order in $U$, since the first order is the Hartree shift, $\rho U$. 

\indent As our solutions given by Eqs. (\ref{Noltingregained}) 
satisfy the first four sum rules or moments, our Green function 
is beyond the Hubbard-I solution\cite{Izyumov-Sckryabin} 
and as a consequence we have an improved solution\cite{Micnas-et-al}. 
The drawback of the Hubbard-I solution, i.e., a gap for any value of the 
interaction, was pointed out by Laura Roth\cite{LRot} many years 
ago. This gap also exists, for all values of $U$, in the spherical 
approximation of 
Nolting\cite{Nolting,Micnas-et-al,JJSS}. The $B$-term gives rise to 
magnetism, a feature which is not present in the Hubbard-I solution.

        In Fig. 1 we present 
the spin susceptibility, $\chi(T)$ vs 
$T/t$ for the two-pole Ansatz for the 
one-particle Green function, or the one-pole 
Ansatz for the self-energy (See Eq. (\ref{ansatz})), 
for $U/t = 4.0$ at half-filling. We see that the 
spin susceptibility looks Curie-like at large 
temperatures and it has a bending towards zero 
at low temperatures. This type of behavior is 
similar to the results found for the attractive 
Hubbard model at half-filling. Let us remember that 
the attractive and repulsive Hubbard model become 
identical at half-filling.  In Fig. 2 we show the 
band-narrow parameter, $B(T)$ vs $T/t$ 
for the same parameters of Fig. 1. We comment here that the 
band-narrowing factor, $B(T)$, in general, 
is $\vec{k}$-dependent. This dependence has been explicitly been shown 
recently to be important by Eskes and Ole\'s\cite{Oles}.

	In summary we have proposed a self-energy of one 
pole which implies that the Green function is composed of two poles. 
This has been accomplished by the use of Dyson's equation 
(Eq. (\ref{1PGF})). We have reproduced with minor effort the Nolting's 
solution\cite{Nolting}-\cite{Micnas-et-al}. The one-pole 
self-energy leading to Nolting's solution puts on 
firm grounds the results found in Refs.\cite{Micnas-et-al} 
where it was argued that the two branches of the one-particle 
Green function were due to a single pole in the 
self-energy. With a single pole-ansatz for the self-energy, we recuperate known 
results in the literature with less analytical effort, at the same time sheding 
light on the structure of the self-energy itself. Our Green's function 
goes beyond the Hubbard-I 
solution because we have imposed the condition 
that the third and four moments be satisfied. For us, Hubbard-I 
means that the first two moments are satisfied.  
We add that besides these features we have justified that, 
for the particular pole structure for our self-energy, this one represents an  expansion in powers of $U$.  The first order is represented 
by the Hartree shift, or $\rho U$. (See Eq. 
(\ref{ansatz})).  With our rather simple approach we have 
been able to have a Green's function with two  
poles. Our approch is different from the one of 
the limit of infinite dimensions\cite{Vollhardt} 
where only the dynamical properties are taken into 
account leaving aside the study of the long range behavior. The moment 
approach is a reliable tool to study strongly correlated 
electronic systems, in particular, the Hubbard model. A 
recent calculation by Nolting, Jaya and Rex\cite{NJR} has applied it to the 
periodic Anderson model, where the relevant quantity of study is the 
self-energy. 
\begin{center}
{\Large Acknowledments}
\end{center}
We would like to thank the CNPq (project No.300705/95-6)
and also from CONICIT (project F-139).We thank Mar\'{\i}a Dolores Garc\'{\i}a
for reading the manuscript. Fruitful discussions with Prof. M.A. 
Continentino, Prof. E. Anda, Prof. M.S. Figueira, Dr. M.H. Pedersen and 
Prof. H. Beck are fully appreciated.\\

\vspace{1.6cm}

\begin{center}
{\Large Figures}\\
\end{center}
\vspace{0.8cm}
\noindent Fig. 1. The static spin susceptibily, 
$\chi(T)$, vs $T/t$ for the one-pole Ansatz for 
the self-energy. Here $U/t = 4.0$ and half-filling. The expression 
for $\chi(T)$ has been borrowed from Ref.\cite{Micnas-et-al}.\\ 

\vspace{0.4cm}

\noindent Fig. 2. The narrowing band parameter, 
$B(T)$ vs $T/t$, for the same parameters of Fig. 1.\\
 
\end{document}